# Micro -Thermonuclear AB-Reactors for Aerospace*

**Alexander Bolonkin**
C&R, 1310 Avenue R, #F-6, Brooklyn, NY 11229, USA
T/F 718-339-4563, aBolonkin@juno.com, aBolonkin@gmail.com, http://Bolonkin.narod.ru

**Abstract**

About fifty years ago, scientists conducted R&D of a thermonuclear reactor that promises a true revolution in the energy industry and, especially, in aerospace.  Using such a reactor, aircraft could undertake flights of very long distance and for extended periods and that, of course, decreases a significant cost of aerial transportation, allowing the saving of ever-more expensive imported oil-based fuels.  (As of mid-2006, the USA's DoD has a program to make aircraft fuel from domestic natural gas sources.)  The temperature and pressure required for any particular fuel to fuse is known as the Lawson criterion $L$. Lawson criterion relates to plasma production temperature, plasma density and time. The thermonuclear reaction is realised when $L > 10^{14}$.  There are two main methods of nuclear fusion: inertial confinement fusion (ICF) and magnetic confinement fusion (MCF).  Existing thermonuclear reactors are very complex, expensive, large, and heavy. They cannot achieve the Lawson criterion.

The author offers several innovations that he first suggested publicly early in 1983 for the AB multi-reflex engine, space propulsion, getting energy from plasma, etc. (see: A. Bolonkin, Non-Rocket Space Launch and Flight, Elsevier, London, 2006, Chapters 12, 3A). It is the micro-thermonuclear AB-Reactors. That is new micro-thermonuclear reactor with very small fuel pellet that uses plasma confinement generated by multi-reflection of laser beam or its own magnetic field. The Lawson criterion increases by hundreds of times. The author also suggests a new method of heating the power-making fuel pellet by outer electric current as well as new direct method of transformation of ion kinetic energy into harvestable electricity.  These offered innovations dramatically decrease the size, weight and cost of thermonuclear reactor, installation, propulsion system and electric generator. Non-industrial countries can produce these researches and constructions.  Currently, the author is researching the efficiency of these innovations for two types of the micro-thermonuclear reactors: multi-reflection reactor (ICF) and self-magnetic reactor (MCF).

**Key words:** Micro-thermonuclear reactor, Multi-reflex AB-thermonuclear reactor, Self-magnetic AB-thermonuclear reactor, aerospace thermonuclear engine.

*Presented as paper AIAA-2006-8104 in 14th Space Plane and Hypersonic Systems Conference, 6-8 November, 2006, USA.

## Introduction
### Brief information about thermonuclear reactors

**Fusion power** is useful energy generated by nuclear fusion reactions. In this kind of reaction two light atomic nuclei fuse together to form a heavier nucleus and release energy. The largest current experiment, JET, has resulted in fusion power production somewhat larger than the power put into the plasma, maintained for a few seconds. In June 2005, the construction of the experimental reactor ITER, designed to produce several times more fusion power than the power into the plasma over many minutes, was announced. The production of net electrical power from fusion is planned for the next generation experiment after ITER.

Unfortunately, this task is not easy, as scientists thought early. Fusion reactions require a very large amount of energy to initiate in order to overcome the so-called *Coulomb barrier* or *fusion barrier energy*. The key to practical fusion power is to select a fuel that requires the minimum amount of energy to start, that is, the lowest barrier energy. The best fuel from this standpoint is a one-to-one mix of deuterium and tritium; both are heavy isotopes of hydrogen. The D-T (Deuterium & Tritium) mix has a low barrier energy. In order to create the required conditions, the fuel must be heated to tens of millions of degrees, and/or compressed to immense pressures.

At present, D-T is used by two main methods of fusion: inertial confinement fusion (ICF) and magnetic confinement fusion (MCF)(for example, tokamak).

In **inertial confinement fusion (ICF)**, nuclear fusion reactions are initiated by heating and compressing a target. The target is a pellet that most often contains deuterium and tritium (often only micro or milligrams). Intense laser or ion beams are used for compression. The beams explosively detonate the outer layers of the target. That accelerats the underlying target layers inward, sending a shockwave into the center of pellet mass. If the shockwave is powerful enough and if high enough density at the center is achieved some of the fuel will be heated enough to cause fusion reactions. In a target which has been heated and compressed to the point of thermonuclear ignition, energy can then heat surrounding fuel to cause it to fuse as well, potentially releasing tremendous amounts of energy.

Fusion reactions require a very large amount of energy to initiate in order to overcome the so-called *Coulomb barrier* or *fusion barrier energy*.

**Magnetic confinement fusion (MCF).** Since plasmas are very good electrical conductors, magnetic fields can also confine fusion fuel. A variety of magnetic configurations can be used, the basic distinction being between magnetic mirror confinement and toroidal confinement, especially tokamaks and stellarators.

**Lawson criterion**. In nuclear fusion research, the Lawson criterion, first derived by John D. Lawson in 1957, is an important general measure of a system that defines the conditions needed for a fusion reactor to reach **ignition**, that is, that the heating of the plasma by the products of the fusion reactions is sufficient to maintain the temperature of the plasma against all losses without external power input. As originally formulated the Lawson criterion gives a minimum required value for the product of the plasma (electron) density $n_e$ and the "energy confinement time" $\tau$. Later analyses suggested that a more useful figure of merit is the "triple product" of density, confinement time, and plasma temperature $T$. The triple product also has a minimum required value, and the name "Lawson criterion" often refers to this inequality.

The key to practical fusion power is to select a fuel that requires the minimum amount of energy to start, that is, the lowest barrier energy. The best fuel from this standpoint is a one-to-one mix of deuterium and tritium; both are heavy isotopes of hydrogen. The D-T (Deuterium & Tritium) mix has a low barrier.

In order to create the required conditions, the fuel must be heated to tens of millions of degrees, and/or compressed to immense pressures. The temperature and pressure required for any particular fuel to fuse is known as the Lawson criterion. For the D-T reaction, the physical value is about

$$L = n_e T \tau > (10^{14} \div 10^{15}) \quad \text{in} \quad \text{"cgs"} \quad \text{units}$$
$$\text{or} \quad L = n T \tau > (10^{20} \div 10^{21}) \quad \text{in} \quad \text{CI} \quad \text{units} \quad ,$$

where $T$ is temperature, [KeV], 1 eV = $1.16 \times 10^4$ °K; $n_e$ is matter density, [1/cm$^3$]; $n$ is matter density, [1/m$^3$]; $\tau$ is time, [s]. Last equation is in metric system. The thermonucler rection of $^2$H + $^3$D realises if $L > 10^{20}$ in CI (meter, kilogram, second) units or $L > 10^{14}$ in 'cgs' (cantimetr, gram, second) units.

This number has not yet been achieved in any reactor, although the latest generations of machines have come close. For instance, the reactor TFTR has achieved the densities and energy lifetimes needed to achieve Lawson at the temperatures it can create, but it cannot create those temperatures at the same time. Future ITER aims to do both.

The Lawson criterion applies to inertial confinement fusion as well as to magnetic confinement fusion but is more usefully expressed in a different form. Whereas the energy confinement time in a magnetic system is very difficult to predict or even to establish empirically, in an inertial system it must be on the order of the time it takes sound waves to travel across the plasma:

$$\tau \approx \frac{R}{\sqrt{kT/m_i}}$$

where $\tau$ is time, s; $R$ is distance, m; $k$ is Boltzmann constant; $m_i$ is mass of ion, kg.

Following the above derivation of the limit on $n_e\tau_E$, we see that the product of the density and the radius must be greater than a value related to the minimum of $T^{3/2}/<\sigma v>$ (here $\sigma$ is Boltzmann constant, v is ion speed). This condition is traditionally expressed in terms of the mass density $\rho$: $\rho R > 1$ g/cm² .

To satisfy this criterion at the density of solid D+T (0.2 g/cm³) would require an implausibly large laser pulse energy. Assuming the energy required scales with the mass of the fusion plasma ($E_{laser} \sim \rho R^3 \sim \rho^{-2}$), compressing the fuel to $10^3$ or $10^4$ times solid density would reduce the energy required by a factor of $10^6$ or $10^8$, bringing it into a realistic range. With a compression by $10^3$, the compressed density will be 200 g/cm³, and the compressed radius can be as small as 0.05 mm. The radius of the fuel before compression would be 0.5 mm. The initial pellet will be perhaps twice as large since most of the mass will be ablated during the compression.

The fusion power density is a good figure of merit to determine the optimum temperature for magnetic confinement, but for inertial confinement the fractional burn-up of the fuel is probably more useful. The burn-up should be proportional to the specific reaction rate ($n^2<\sigma v>$) times the confinement time (which scales as $T^{1/2}$) divided by the particle density $n$:

$$\text{burn-up fraction} \sim n^2<\sigma v> T^{1/2} / n \sim (nT) (<\sigma v>/T^{3/2})$$

Thus the optimum temperature for inertial confinement fusion is that which maximizes $<\sigma v>/T^{3/2}$, which is slightly higher than the optimum temperature for magnetic confinement.

**Short history of thermonuclear fusion**. One of the earliest (in the late 1970's and early 1980's) serious attempts at an ICF design was **Shiva**, a 20-armed neodymium laser system built at the Lawrence Livermore National Laboratory (LLNL) that started operation in 1978. Shiva was a "proof of concept" design, followed by the **NOVA** design with 10 times the power. Funding for fusion research was severely constrained in the 80's, but NOVA nevertheless successfully gathered enough information for a next generation machine whose goal was ignition. Although net energy can be released even without ignition (the breakeven point), ignition is considered necessary for a *practical* power system.

The resulting design, now known as the National Ignition Facility, commenced being constructed at LLNL in 1997. Originally intended to start construction in the early 1990s, the NIF is now six years behind schedule and overbudget by over $1.4 billion. Nevertheless many of the problems appear to be due to the "big lab" mentality and shifting the focus from pure ICF research to the nuclear stewardship program, LLNLs traditional nuclear weapons-making role. NIF is now scheduled to "burn" in 2010, when the remaining lasers in the 192-beam array are finally installed.

Laser physicists in Europe have put forward plans to build a £500m facility, called HiPER, to study a new approach to laser fusion. A panel of scientists from seven European Union countries believes that a "fast ignition" laser facility could make a significant contribution to fusion research, as well as supporting experiments in other areas of physics. The facility would be designed to achieve high energy gains, providing the critical intermediate step between ignition and a demonstration reactor. It would consist of a long-pulse laser with an energy of 200 kJ to compress the fuel and a short-pulse laser with an energy of 70 kJ to heat it.

Confinement refers to all the conditions necessary to keep a plasma dense and hot long enough to undergo fusion:

- **Equilibrium:** There must be no net forces on any part of the plasma, otherwise it will rapidly disassemble. The exception, of course, is inertial confinement, where the relevant physics must occur faster than the disassembly time.
- **Stability:** The plasma must be so constructed that small deviations are restored to the initial state, otherwise some unavoidable disturbance will occur and grow exponentially until the plasma is destroyed.
- **Transport:** The loss of particles and heat in all channels must be sufficiently slow. The word "confinement" is often used in the restricted sense of "energy confinement".

To produce self-sustaining fusion, the energy released by the reaction (or at least a fraction of it) must be used to heat new reactant nuclei and keep them hot long enough that they also undergo fusion reactions. Retaining the heat generated is called energy **confinement** and may be accomplished in a number of ways.

The hydrogen bomb weapon has no confinement at all. The fuel is simply allowed to fly apart, but it takes a certain length of time to do this, and during this time fusion can occur. This approach is called **inertial confinement** (fig.1). If more than about a milligram of fuel is used, the explosion would destroy the machine, so controlled thermonuclear fusion using inertial confinement causes tiny pellets of fuel to explode several times a second. To induce the explosion, the pellet must be compressed to about 30 times solid density with energetic beams. If the beams are focused directly on the pellet, it is called **direct drive**, which can in principle be very efficient, but in practice it is difficult to obtain the needed uniformity. An alternative approach is **indirect drive**, in which the beams heat a shell, and the shell radiates x-rays, which then implode the pellet. The beams are commonly laser beams, but heavy and light ion beams and electron beams have all been investigated and tried to one degree or another.

They rely on fuel pellets with a "perfect" shape in order to generate a symmetrical inward shock wave to produce the high-density plasma, and in practice these have proven difficult to produce. A recent development in the field of laser-induced ICF is the use of ultra-short pulse multi-petawatt lasers to heat the plasma of an imploding pellet at exactly the moment of greatest density after it is imploded conventionally using terawatt scale lasers. This research will be carried out on the (currently being built) OMEGA EP petawatt and OMEGA lasers at the University of Rochester and at the GEKKO XII laser at the Institute for Laser Engineering in Osaka Japan which, if fruitful, may have the effect of greatly reducing the cost of a laser fusion-based power source.

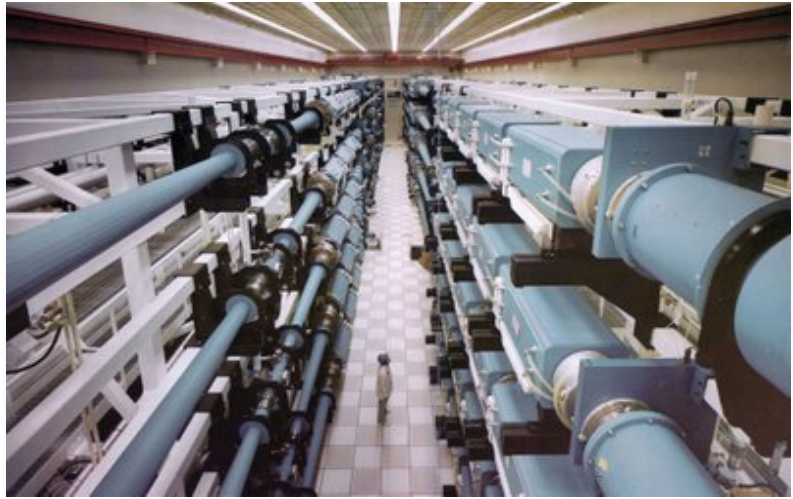

**Fig.1**. Laser installation for NOVA inertial thermonuclear reactor. Look your attention in the man and gigantic size of laser installation for reactor. Cost is some billions of dollars.

At the temperatures required for fusion, the fuel is in the form of a plasma with very good electrical conductivity. This opens the possibility to confine the fuel and the energy with magnetic fields, an idea known as **magnetic confinement** (fig.2).

Much of this progress has been achieved with a particular emphasis on tokamaks (fig.2).

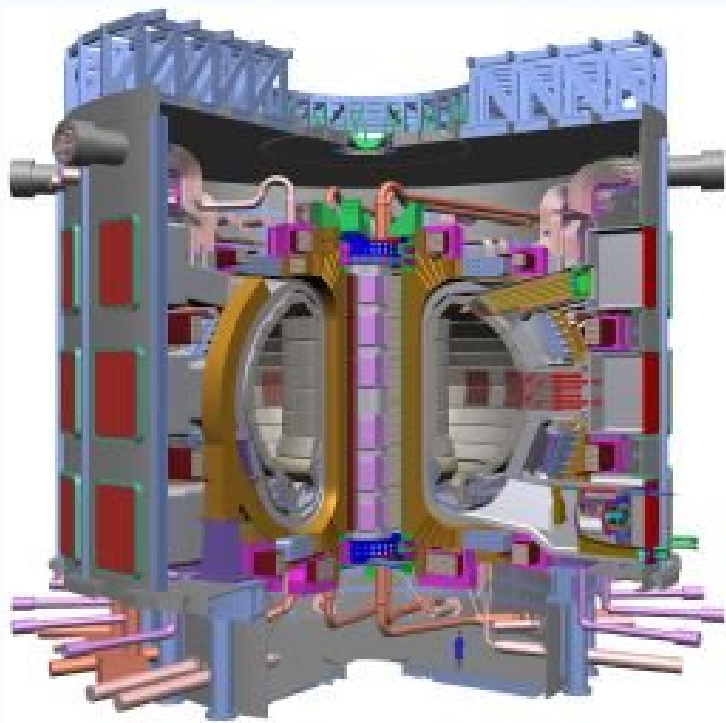

**Fig. 2**. Magnetic thermonuclear reactor. The size of the installation is obvious if you compare it with the "Little Blue Man" inside the machine at the bottom. Cost is some tens of billions of dollars.

In fusion research, achieving a fusion energy gain factor $Q = 1$ is called **breakeven** and is considered a significant although somewhat artificial milestone. **Ignition** refers to an infinite $Q$, that is, a self-sustaining plasma where the losses are made up for by fusion power without any external input. In a practical fusion reactor, some external power will always be required for things like current drive, refueling, profile control, and burn control. A value on the order of $Q = 20$ will be required if the plant is to deliver much more energy than it uses internally.

In a fusion power plant, the nuclear island has a **plasma chamber** with an associated vacuum system, surrounded by a plasma-facing components (first wall and divertor) maintaining the vacuum boundary and absorbing the thermal radiation coming from the plasma, surrounded in turn by a blanket where the neutrons are absorbed to breed tritium and heat a working fluid that transfers the power to the balance of plant. If magnetic confinement is used, a **magnet** system, using primarily cryogenic superconducting magnets, is needed, and usually systems for heating and refueling the plasma and for driving current. In inertial confinement, a **driver** (laser or accelerator) and a focusing system are needed, as well as a means for forming and positioning the **pellets**.

The magnetic fusion energy (MFE) program seeks to establish the conditions to sustain a nuclear fusion reaction in a plasma that is contained by magnetic fields to allow the successful production of fusion power.

In thirty years, scientists have increased the Lawson criterion of the ICF and tokamak installations by tens of times. Unfortunately, all current and some new installations (ICF and totamak) have a Lawrence criterion that is tens of times lower than is necessary (fig.3).

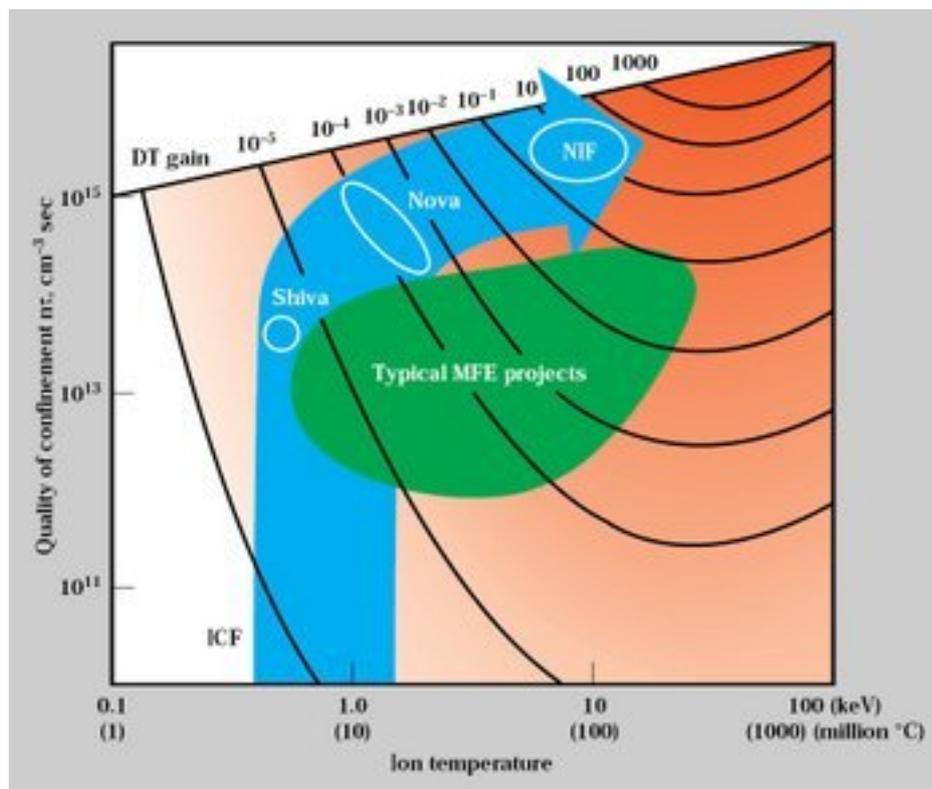

**Fig. 3.** Parameter space occupied by inertial fusion energy and magnetic fusion energy devices. The regime allowing thermonuclear ignition with high gain lies near the upper right corner of the plot.

# Innovation

As you can see in the Equation for thermonuclear reaction (reaction's "ignition") it is necessary to rapidly and greatly increase the target-enveloping temperature, the density of target proper and to shorten the time of the operation in order to keep the fuel in these precisely induced conditions. In ICF the density of plasma is very high ($10^{28}$ m$^{-3}$, it increases in 20-30 times in target), the temperature reaches tens of millions °K, but time is measured in nanoseconds. As a result, the Lawson criterion is tens to hundreds of times lower than is required. In a tokomak, the time is mere parts of second and the ambient temperature is tens of millions of degrees, but density of plasma is very small ($10^{20}$ m$^{-3}$). The Lawson criterion is also tens to hundreds of times lower than needed.

The author offers some innovations and names these reactors as AB-reactors.. The main innovations are below.

**Multi-reflect reactor (MRR)**. The first innovation suggested early in 1983 [14] and developed later in [1]-[26] for multi-reflex engine and space propulsion. Conventional ICF has conventional inside surface of the combustion chamber. This surface absorbs part of the heat radiation emanating from the pellet and plasma, the rest of the radiation reflects in all directions and is also absorbed by walls of combustion chamber. As result the target loses energy expensively delivered by lasers. This loss is so huge that we need very powerful lasers and we cannot efficiently heat the target to reach ignition temperature (Lawson's criterion). In all current ICF installations this loss is tens of times more than is acceptable.

The innovative ICF has, on the inside surface of combustion chamber, a covering of small Prism Reflectors (PR) (figs.4,5) (or multi-layer reflector. Note: Multi-layer reflector can only reflect the laser beam). The system of prism reflectors has great advantages in comparison with conventional mirror and especially with conventional combustion chamber. The advantages are listed:

1. The prism reflector has very high efficiency. The coefficient of its radiate absorption is less about million times the rate of the conventional mirror.
2. The prism mirror reflects the radiation in widely diapason of continuous spectrum. A conventional mirror reflects the radiation only in narrow diapason of continuous spectrum. That means that any conventional mirror has big absorption of radiation energy even if it has high reflectivity (up 99%) in narrow interval of the continuous spectrum. The prism reflector allows to use the thermal plasma radiation.
3. The prism reflector bounces the heat radiation exactly to a point where heat beam comes up, even if it has defect at position. The conventional mirror having small defect in position (or the pellet is not located exactly in center of sphere) destroys the pellet.
4. According with Point 3 above, the prism reflector may be used in cylindrical (toroidal) camera (fig.5) (tokomak, stellarator). A conventional mirror cannot be employed because reflected ray will be sent in the other direction.
5. The prism reflector can uniformly distribute the beam energy in pellet surface. The small spherical plasma pellet reflected the scattered radiation. That means the laser beam after the first reflection reflects on semi-sphere, after two reflections that presses on full sphere, after 3 - 4 reflection the pressure is almost uniformly. That decreases a number of needed laser beams, simplify, and decreases cost of laser installation.

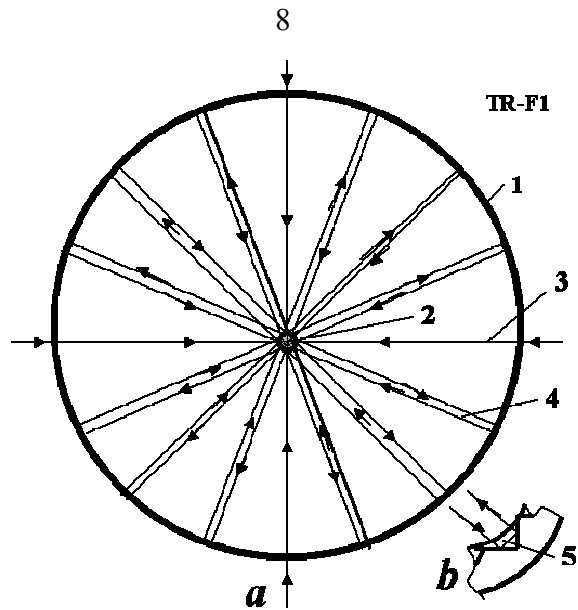

**Fig.4.** Multi-reflex Reactor. (a) Cross-section of ICF; (b) Cross-section of spherical combustion chamber and prism reflectors. Notations: 1 - spherical shell; 2 - target (pellet); 3- ignition laser beams; 4 - reflected laser beams; 5 - prism reflector.

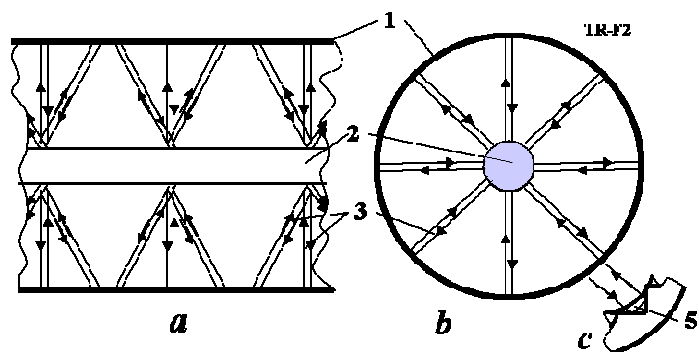

**Fig.5.** Multi-reflective in cylindrical tube chamber or tokomak. (a) Cross-section along long axis. (b) Cross-section along transverse axis; (c) Cross-section of toroidal or cylindrical combustion and prism reflectors. Notations: 1 - combustion chamber; 2 - plasma (fuel capsule); 3 - reflected laser beams; 5 - prism reflectors.

In particular, this innovation may be used in already built current reactors for their improvement.

  **Self-Magnetic reactor (SMR)** (Figs. 6,7). The magnetic pressure is proportional to the inverse value of electric conductor diameter. (The conventional magnetic reactor has a diameter of plasma flex some meters). The high temperature plasma has excellent conductivity which does not depend from plasma density. If the diameter is decreased to 0.1 mm and electric currency is high, the magnetic pressure is increased by hundreds or thousand times and that can keep the high-density plasma. Thus, the plasma is confined by self-generated magnetic field (and by pinch-effect) and it does not need in powerful outer magnetic field created very complex, high cost super-conductivity system! This innovation in MCF is dramatically decreasing the size of reaction zone and using of gaseous compression fuel pellet (micro-capsule) in magnetic confinement reactor.
   The other innovation in SMR is uses the electric current (electric impulse) for initial heating of microcapsule targets. That means we don't need a large, very complex and expensive laser (or ion beam) system for inertial confinement reactor or induce system in magnetic confinement reactor. That is possible in special design of the fuel microcapsule. The energy for heating of the microcapsule to thermonuclear temperature is small and conventional condensers may be used for it.

The some innovations are magnetic and radiation stoppers (confinements) of plasma in ends of the fuel capsule. It is suggested two methods:

1) The capsule has conic ends (fig. 7d). The capsule radius decreases at ends. That means the magnetic intensity will increase at the capsule ends (up 10 and more times). They will work as magnetic mirror (plasma stopper) (fig. 7).
2) Our magnetic field has other direction and form then it is in conventional tokomak. In tokomak the magnetic lines are parallel to a toroid (or cylinder) axis. In our SMR the magnetic lines are circles around cylindrical capsule. That means there is an axis magnetic force which put obstacles a heat transfer from plasma to electric electrodes. That works as radial and axial magnetic stopper.

There are others innovations which reader can apprehend in comparison the offered micro AB-reactors and current and under construction reactors.

These innovations decrease the size, weight and the monetary cost of thermonuclear reactors by thousands of times and allow the widespread future construction of thermonuclear electric stations, engines, and space propulsions.

The offered self-magnetic reactor has the following differences in comparison with Z-machine of Sandia Laboratory (USA). Z-machine used a set of very fine tungsten wires running around the fuel would be "dumped" with the current instead. The wires would quickly vaporize into a plasma, which is conductive, and the current flow would then cause the plasma to pinch. The key difference is that the plasma would not be the fuel, as in our SMCF, but used solely to generate very high-energy X-rays as the metal plasma compressed and heated. The x-rays would compress a tiny fuel cylinder containing deuterium-tritium mix, in the same fashion that the X-rays generated from a nuclear bomb compress the fuel load in an H-bomb. The superpower x-ray output pulse (up 2.7 megajouls!) generated by heavy tungsten metal plasma ($_{184}^{74}W$) is very danger for people. In additional, the powerful fluctuation in the magnetic field (an "electromagnetic pulse") also generates strong electric current in all of the metallic objects in the room and demiges electronic devices.

In our machime the small fuel cylinder has a thin conductive layer from light metall. The capsule can also have contactivity filaments into fuel. They help to produce initial heating of fuel plasma (up $10^4 \div 10^5$ °K) and initial the plasma compression (rocket and/or inertial). The father plasma heating and confinement is produced by voltage curve of fig.17 which create self-magnetic field equil (or more) plasma gas pressure.

**Summary.** This work offers two types of micro-AB-thermonuclear reactors: by multi-reflex radiation confinement of plasma and self-magnetic confinement. They can use high and low density fuel (compressed gas or liquid/frizzed fuel) and they can work in pulse or continuous regimes.

The offered micro-AB-reactor with self-magnetic confinement includes: micro fuel capsule with compressed gaseous or liquid (frizzed) fuel; two electric electrodes, and a combustion chamber. Internal surface of combustion chamber is covered by prism or multi-layer reflectors.

The capsule contains thermonuclear fuel (it conventionally has two component, example D + T), and conducting capsule shell. Fuel may be composed by conducting fiber for quick heating. The capsule has the conic ends.

The electric electrodes have windings for creating magnetic locks, canals for feeding of fuel capsule (or injector for liquid fuel), and electron injector (electric currency). Last may be electron (currency) emitter. Electrodes also contain a cooling system and thermo-protection.

The suggested reactor works the following way. The strong impulse electric current passes through capsule. The capsule shell explodes, creating an initial plasma flux and compressed, heating, and creating initial fuel plasma fuse. The plasma radiation erupts and part of them returns and compresses the plasma, helping the electric current to heat the plasma to its ignition temperature.

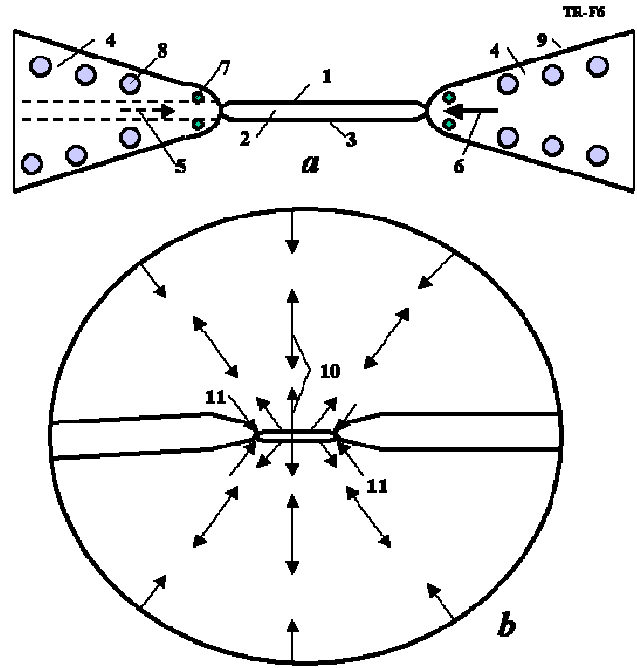

**Fig. 6.** Micro AB-reactor with self-magnetic confinement and radiation support of plasma. (a) – fuel micro-capsule and electrodes; (b) - Reflective camera. Notation: 1 - micro fuel capsule; 2 - thermonuclear fuel into capsule; 3 - capsule shell and covering; 4 - electrodes; 5 - feeding of capsule; 6 - electric currency (electron injector); 7 - magnetic stopper; 8 - cooling system; 9 - thermo protection; 10 - radiation; 11 - additional radiation pressure to fuel capsule ends.

The self-magnetic reactor uses very small capsule diameter when the magnetic intensity is very high. The magnetic intensity has good distribution (decreases to plasma center, fig. 7c) and magnetic pressure is big (it is enough to keep the kinetic plasma pressure which is not so much for low density plasma).

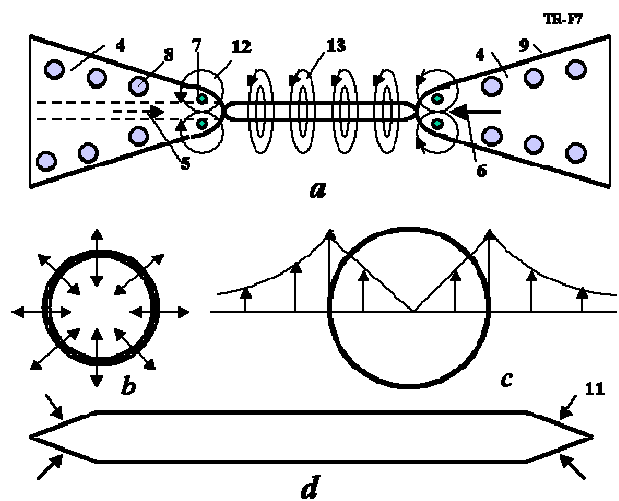

**Fig.7.** Micro AB-thermonuclear reactor with self-magnetic confinement. (a) Self-magnetic field around fuel capsule, (b) - Explosion initial compression, (c) Magnetic intensity distribution in cross-section plasma flex, (d) Fuel capsule. Notations: 4 - electrodes; 5 - feeding of capsule; 6 - electric currency (electron injector); 7

- magnetic stopper; 8 - cooling system; 9 - thermo protection; 11 - additional end capsule pressure of self-magnetic field and radiation; 12 - magnetic corks. 13 - self-magnetic field of capsule (and pinch-effect),

We spoke about micro AB-reactors. But it does not mean that power of them is small. In next articles I will discuss the methods for transformation and utilization of the thermonuclear energy into other types of energy and propulsion. These completed research show the power of micro AB-reactor can reach some thousands of kW.

The computations of the offered reactors are presented below.

# Theory of current thermonuclear reactor

## Methods of confinement in current reactors

**Magnetic confinement**. Magnetic fields can confine fusion fuel because plasma is a very good electrical conductor. A variety of magnetic configurations can be used, the most basic distinction being tokamaks and stellarators.

**Inertial confinement.** A third confinement principle is to apply a rapid pulse of energy to a large part of the surface of a pellet of fusion fuel, causing it to simultaneously "implode" and heat to very high pressure and temperature. If the fuel is dense enough and hot enough, the fusion reaction rate will be high enough to burn a significant fraction of the fuel before it has dissipated. To achieve these extreme conditions, the initially cold fuel must be explosively compressed. Inertial confinement is used in the hydrogen bomb, where the driver is x-rays created by a fission bomb. Inertial confinement is also attempted in "controlled" nuclear fusion, where the driver is a laser, ion, or electron beam.

Some other confinement principles have been investigated, such as muon-catalyzed fusion, the Farnsworth-Hirsch fusor (inertial electrostatic confinement), and bubble fusion.

In man-made fusion, the primary fuel is not constrained to be protons and higher temperatures can be used, so reactions with larger cross-sections are chosen. This implies a lower Lawson criterion, and therefore less startup effort. Another concern is the production of neutrons, which activate the reactor structure radiologically, but also have the advantages of allowing volumetric extraction of the fusion energy and tritium breeding. Reactions that release no neutrons are referred to as *aneutronic*.

In order to be useful as a source of energy, a fusion reaction must satisfy several criteria. It must:

- **be exothermic** - This may be obvious, but it limits the reactants to the low Z (number of protons) side of the curve of binding energy. It also makes helium $^4$He the most common product because of its extraordinarily tight binding, although $^3$He and $^3$H also show up.
- **involve low Z nuclei** - This is because the electrostatic repulsion must be overcome before the nuclei are close enough to fuse.
- **have two reactants** - At anything less than stellar densities, three body collisions are too improbable. It should be noted that in inertial confinement, both stellar densities and temperatures are exceeded to compensate for the shortcomings of the third parameter of the Lawson criterion, ICF's very short confinement time.
- **have two or more products** - This allows simultaneous conservation of energy and momentum without relying on the (weak!) electromagnetic force.
- **conserve both protons and neutrons** - The cross sections for the weak interaction are too small.

Few reactions meet these criteria. The following are those with the largest cross-sections:

**Table 1.** Sutable reactions for thermonuclear fusion.

| (1)    | D      | + | T      | →   | $^4$He (3.5 MeV)  | + |   | n         | (14.1 MeV) |   |              |     |
|--------|--------|---|--------|-----|-------------------|---|---|-----------|------------|---|--------------|-----|
| (2i)   | D      | + | D      | →   | T (1.01 MeV)      | + |   | p         | (3.02 MeV) |   |              | *50%* |
| (2ii)  |        |   |        | →   | $^3$He (0.82 MeV) | + |   | n         | (2.45 MeV) |   |              | *50%* |
| (3)    | D      | + | $^3$He | →   | $^4$He (3.6 MeV)  | + |   | p         | (14.7 MeV) |   |              |     |
| (4)    | T      | + | T      | →   | $^4$He            |   | +2 | n        | + 11.3 MeV |   |              |     |
| (5)    | $^3$He | + | $^3$He | →   | $^4$He            |   | +2 | p        | + 12.9 MeV |   |              |     |
| (6i)   | $^3$He | + | T      | →   | $^4$He            | + |   | p         |            | + n | + 12.1 MeV | *51%* |
| (6ii)  |        |   |        | →   | $^4$He (4.8 MeV)  | + |   | D         | (9.5 MeV)  |   |              | *43%* |
| (6iii) |        |   |        | →   | $^4$He (0.5 MeV)  | + |   | n         | (1.9 MeV)  | + p | (11.9 MeV) | *6%* |
| (7)    | D      | + | $^6$Li | → 2 | $^4$He + 22.4 MeV |   |   |           |            |   |              |     |
| (8)    | p      | + | $^6$Li | →   | $^4$He (1.7 MeV)  | + |   | $^3$He    | (2.3 MeV)  |   |              |     |
| (9)    | $^3$He | + | $^6$Li | → 2 | $^4$He            |   | + | p         | + 16.9 MeV |   |              |     |
| (10)   | p      | + | $^{11}$B | → 3 | $^4$He + 8.7 MeV |   |   |           |            |   |              |     |

p *(protium),* D *(deuterium), and* T *(tritium) are shorthand notation for the main three isotopes of hydrogen.*

For reactions with two products, the energy is divided between them in inverse proportion to their masses, as shown. In most reactions with three products, the distribution of energy varies. For reactions that can result in more than one set of products, the branching ratios are given.

Some reaction candidates can be eliminated at once. The D-$^6$Li reaction has no advantage compared to p-$^{11}$B because it is roughly as difficult to burn but produces substantially more neutrons through D-D side reactions. There is also a p-$^7$Li reaction, but the cross-section is far too low except possible for $T_i$ > 1 MeV, but at such high temperatures an endothermic, direct neutron-producing reaction also becomes very significant. Finally there is also a p-$^9$Be reaction, which is not only difficult to burn, but $^9$Be can be easily induced to split into two alphas and a neutron.

In addition to the fusion reactions, the following reactions with neutrons are important in order to "breed" tritium in "dry" fusion bombs and some proposed fusion reactors:

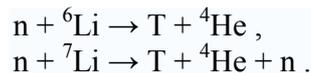

$$n + {}^6\text{Li} \rightarrow T + {}^4\text{He} ,$$
$$n + {}^7\text{Li} \rightarrow T + {}^4\text{He} + n .$$

To evaluate the usefulness of these reactions, in addition to the reactants, the products, and the energy released, one needs to know something about the cross section. Any given fusion device will have a maximum plasma pressure that it can sustain, and an economical device will always operate near this maximum. Given this pressure, the largest fusion output is obtained when the temperature is selected so that $\langle\sigma v\rangle/T^2$ is a maximum. This is also the temperature at which the value of the triple product $nT\tau$ required for ignition is a minimum. This chosen optimum temperature and the value of <σv>/T² at that temperature is given for a few of these reactions in the following table.

**Table 2**. Optimum temperature and the value of <σv>/T² at that temperature

| fuel | $T$ [keV] | <σv>/T² [m³/s/keV²] |
|---|---|---|
| D-T | 13.6 | $1.24 \times 10^{-24}$ |
| D-D | 15 | $1.28 \times 10^{-26}$ |
| D-$^3$He | 58 | $2.24 \times 10^{-26}$ |
| p-$^6$Li | 66 | $1.46 \times 10^{-27}$ |
| p-$^{11}$B | 123 | $3.01 \times 10^{-27}$ |

Note that many of the reactions form chains. For instance, a reactor fueled with T and $^3$He will create some D, which is then possible to use in the D + $^3$He reaction if the energies are "right". An elegant idea is to combine the reactions (8) and (9). The $^3$He from reaction (8) can react with $^6$Li in reaction (9) before completely thermalizing. This produces an energetic proton which in turn undergoes reaction (8) before thermalizing. A detailed analysis shows that this idea will not really work well, but it is a good example of a case where the usual assumption of a Maxwellian plasma is not appropriate.

Any of the reactions above can, in principle, be the basis of fusion power production. In addition to the temperature and cross section discussed above, we must consider the total energy of the fusion products $E_{fus}$, the energy of the charged fusion products $E_{ch}$, and the atomic number $Z$ of the non-hydrogenic reactant.

Specification of the D-D reaction entails some difficulties, though. To begin with, one must average over the two branches (2) and (3). More difficult is to decide how to treat the T and $^3$He products. T burns so well in a deuterium plasma that it is almost impossible to extract from the plasma. The D-$^3$He reaction is optimized at a much higher temperature, so the burn-up at the optimum D-D temperature may be low, so it seems reasonable to assume the T but not the $^3$He gets burned up and adds its energy to the net reaction. Thus we will count the D-D fusion energy as $E_{fus}$ = (4.03+17.6+3.27)/2 = 12.5 MeV and the energy in charged particles as $E_{ch}$ = (4.03+3.5+0.82)/2 = 4.2 MeV.

Another unique aspect of the D-D reaction is that there is only one reactant, which must be taken into account when calculating the reaction rate.

With this choice, we tabulate parameters for four of the most important reactions.

**Table 3.** Parameters of the most important reactions.

| Fuel | $Z$ | $E_{fus}$ [MeV] | $E_{ch}$ [MeV] | neutronicity |
|---|---|---|---|---|
| D-T | 1 | 17.6 | 3.5 | 0.80 |
| D-D | 1 | 12.5 | 4.2 | 0.66 |
| D-$^3$He | 2 | 18.3 | 18.3 | ~0.05 |
| p-$^{11}$B | 5 | 8.7 | 8.7 | ~0.001 |

The last column is the **neutronicity** of the reaction, the fraction of the fusion energy released as neutrons. This is an important indicator of the magnitude of the problems associated with neutrons like radiation damage, biological shielding, remote handling, and safety. For the first two reactions it is calculated as ($E_{fus}$-$E_{ch}$)/$E_{fus}$. For the last two reactions, where this calculation would give zero, the values quoted are rough estimates based on side reactions that produce neutrons in a plasma in thermal equilibrium.

Of course, the reactants should also be mixed in the optimal proportions. This is the case when each reactant ion plus its associated electrons accounts for half the pressure. Assuming that the total pressure is fixed, this means that density of the non-hydrogenic ion is smaller than that of the hydrogenic ion by a factor 2/(Z+1). Therefore the rate for these reactions is reduced by the same factor, on top of any differences in the values of <σv>/T². On the other hand, because the D-D reaction has only one reactant, the rate is twice as high as if the fuel were divided between two hydrogenic species.

Thus, there is a "penalty" of (2/(Z+1)) for non-hydrogenic fuels arising from the fact that they require more electrons, which take up pressure without participating in the fusion reaction. There is, at the same time, a "bonus" of a factor 2 for D-D due to the fact that each ion can react with any of the other ions, not just a fraction of them.

We can now compare these reactions in the following table 4.

Table 4. Comparison of reactions

| fuel | <σv>/T² | penalty/bonus | reactivity | Lawson criterion | power density |
|---|---|---|---|---|---|
| D-T | $1.24 \times 10^{-24}$ | 1 | 1 | 1 | 1 |
| D-D | $1.28 \times 10^{-26}$ | 2 | 48 | 30 | 68 |
| D-$^3$He | $2.24 \times 10^{-26}$ | 2/3 | 83 | 16 | 80 |
| p-$^{11}$B | $3.01 \times 10^{-27}$ | 1/3 | 1240 | 500 | 2500 |

The maximum value of <σv>/T² is taken from a previous table. The "penalty/bonus" factor is that related to a non-hydrogenic reactant or a single-species reaction. The values in the column "reactivity" are found by dividing ($1.24 \times 10^{-24}$ by the product of the second and third columns. It indicates the factor by which the other reactions occur more slowly than the D-T reaction under comparable conditions. The column "Lawson criterion" weights these results with $E_{ch}$ and gives an indication of how much more difficult it is to achieve ignition with these reactions, relative to the difficulty for the D-T reaction. The last column is labeled "power density" and weights the practical reactivity with $E_{fus}$. It indicates how much lower the fusion power density of the other reactions is compared to the D-T reaction and can be considered a measure of the economic potential.

## Bremsstrahlung (brake) losses

Bremsstrahlung, (from the German *bremsen*, to brake and *Strahlung*, radiation, thus, "braking radiation"), is electromagnetic radiation produced by the acceleration of a charged particle, such as an electron, when deflected by another charged particle, such as an atomic nucleus. The term is also used to refer to the process of producing the radiation. Bremsstrahlung has a continuous spectrum. The phenomenon was discovered by Nikola Tesla (1856-1943) during high frequency research he conducted between 1888 and 1897.

Bremsstrahlung may also be referred to as free-free radiation. This refers to the radiation that arises as a result of a charged particle that is free both before and after the deflection (acceleration) that causes the emission. Strictly speaking, bremsstrahlung refers to any radiation due to the acceleration of a charged particle, which includes synchrotron radiation; however, it is frequently used (even when not speaking German) in the more literal and narrow sense of radiation from electrons stopping in matter.

The ions undergoing fusion will essentially never occur alone but will be mixed with electrons that neutralize the ions' electrical charge and form a plasma. The electrons will generally have a temperature comparable to or greater than that of the ions, so they will collide with the ions and emit Bremsstrahlung. The Sun and stars are opaque to Bremsstrahlung, but essentially any terrestrial fusion reactor will be optically thin at relevant wavelengths. Bremsstrahlung is also difficult to reflect and difficult to convert directly to electricity, so the ratio of fusion power produced to Bremsstrahlung radiation lost is an important figure of merit. This ratio is generally maximized at a much higher temperature than that which maximizes the power density (see the previous subsection). The following table shows the rough optimum temperature and the power ratio at that temperature for several reactions.

**Table 5.** Rough optimum temperature and the power ratio of fusion and Bremsstrahlung radiation lost.

| Fuel | $T_i$ (keV) | $P_{fusion}/P_{Bremsstrahlung}$ |
|---|---|---|
| D-T | 50 | 140 |
| D-D | 500 | 2.9 |
| D-$^3$He | 100 | 5.3 |
| $^3$He-$^3$He | 1000 | 0.72 |
| p-$^6$Li | 800 | 0.21 |
| p-$^{11}$B | 300 | 0.57 |

The actual ratios of fusion to Bremsstrahlung power will likely be significantly lower for several reasons. For one, the calculation assumes that the energy of the fusion products is transmitted completely to the fuel ions, which then lose energy to the electrons by collisions, which in turn lose energy by Bremsstrahlung. However because the fusion products move much faster than the fuel ions, they will give up a significant fraction of their energy directly to the electrons. Secondly, the plasma is assumed to be composed purely of fuel ions. In practice, there will be a significant proportion of impurity ions, which will lower the ratio. In particular, the fusion products themselves *must* remain in the plasma until they have given up their energy, and *will* remain some time after that in any proposed confinement scheme. Finally, all channels of energy loss other than Bremsstrahlung have been neglected. The last two factors are related. On theoretical and experimental grounds, particle and energy confinement seem to be closely related. In a confinement scheme that does a good job of retaining energy, fusion products will build up. If the fusion products are efficiently ejected, then energy confinement will be poor, too.

The temperatures maximizing the fusion power compared to the Bremsstrahlung are in every case higher than the temperature that maximizes the power density and minimizes the required value of the fusion triple product (Lawson criterion). This will not change the optimum operating point for D-T very much because the Bremsstrahlung fraction is low, but it will push the other fuels into regimes where the power density relative to D-T is even lower and the required confinement even more difficult to achieve. For D-D and D-$^3$He, Bremsstrahlung losses will be a serious, possibly prohibitive problem. For $^3$He-$^3$He, p-$^6$Li and p-$^{11}$B the Bremsstrahlung losses appear to make a fusion reactor using these fuels impossible.

In a plasma, the free electrons are constantly producing Bremsstrahlung in collisions with the ions. The power density of the Bremsstrahlung radiated is given by

$$P_{Br} = \frac{16\alpha^3 h^2}{\sqrt{3}\, m_e^{3/2}} n_e^2 T_e^{1/2} Z_{eff}$$

$T_e$ is the electron temperature, $\alpha$ is the fine structure constant, $h$ is Planck's constant, and the "effective" ion charge state $Z_{eff}$ is given by an average over the charge states of the ions:

$$Z_{eff} = \Sigma\, (Z^2 n_Z) / n_e$$

This formula is derived in "Basic Principles of Plasmas Physics: A Statistical Approach" by S. Ichimaru, p. 228. It applies for high enough $T_e$ that the electron deBroglie wavelength is longer than the classical Coulomb distance of closest approach. In practical units, this formula gives

$$P_{Br} = (1.69 \times 10^{-32}\ /\text{W cm}^{-3})\ (n_e/\text{cm}^{-3})^2\ (T_e/\text{eV})^{1/2}\ Z_{eff}$$
$$= (5.34 \times 10^{-37}\ /\text{W m}^{-3})\ (n_e/\text{m}^{-3})^2\ (T_e/\text{keV})^{1/2}\ Z_{eff}$$

Where Wcm$^{-3}$, cm$^{-3}$, eV, Wm$^{-3}$, m$^{-3}$, keV are units of correcponding magnitudes. For very high temperatures there are relativistic corrections to this formula, that is, additional terms of order $T_e/m_e c^2$.

# List of main equations.

Below are the main equations for estimation of benefits from the offered innovations.

**1. Energy needed for thermonuclear reaction.**

$$F = k\frac{Q_1 Q_2}{r^2},\quad E = \int_{r_0}^{\infty} F dr,\quad E = \frac{k Z_1 Z_2 e^2}{r_0},\qquad\qquad (1)$$
$$r_0 = (1.2 \div 1.5) \cdot 10^{15} \sqrt{A}$$

where $k = 1.38 \times 10^{-23}$ Boltzmann constant, J/°K; $Z_1, Z_2$ are charge state of 1 and 2 particles respectively; $e = 1{,}6 \times 10^{-19}$ C is charge of electron; $r_o$ is radius of nuclear force, m; $A$ is number of element; $F$ is force, N; $E$ is energy, J; $Q$ is charge of particles.

For example, for reaction H+H (hydrogen, $Z_1 = Z_2 = 1$, $r_o \approx 2 \times 10^{-15}$ m) this energy is $\approx 0.7$ MeV or 0.35 MeV for every particle. The real energy is about 30 times less because part of the particles has more average speed and there is a tunnel effect.

**Energy needed for ignition.** Fig. 8 shows a magnitude $n\tau$ (analog of Lawson criterion) required for ignition.

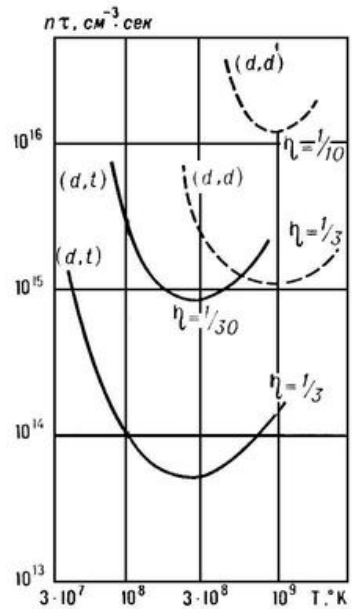

**Fig. 8.** Value $n\tau$ (analog of Lawson criterion) versus temperature in K. Value $n\tau$ is in s/cm$^3$.

At present the industry produces power lasers:
- Carbon dioxide lasers emit up to 100 kW at 9.6 μm and 10.6 μm, and are used in industry for cutting and welding.
- Carbon monoxide lasers must be cooled but can produce up to 500 kW.

Special laser and ICF reastors:
- NOVA (1999, USA). Laser 100 kJ (wavelenght λ=1054×10$^{-9}$ m) and 40 kJ (wavelenght λ=351×10$^{-9}$ m), power few tens of terawatts (1 TW = 10$^{12}$ W), time of impulse $(2 \div 4) \times 10^{-9}$ s, 10-beams, Matter is Nd:class.
- OMERA (1995, USA). 60-beam, neodyminm class laser, 30 kJ, power 60 TW.
- Z-mashine (USA, under constructin), Power is up 350 TW. It can create currency impuls up $20 \times 10^6$ A.
- NIF (USA). As of 2005 the National Ignition Facility is working on a system that, when complete, will contain 192-beam, 1.8-megajoule, 700-terawatt laser system adjoining a 10-meter-diameter target chamber.
- 1.25 PW - world's most powerful laser (claimed on 23 May 1996 by Lawrence Livermore Laboratory).

**3. Radiation energy from hot solid black body** is (Stefan-Boltzmann Law):
$$E = \sigma T^4, \qquad (2)$$
where $E$ is emitted energy, W/m$^2$; $\sigma = 5.67 \times 10^{-8}$ - Stefan-Boltzmann constant, W/m$^2$ °K$^4$; $T$ is temperature in °K.

**4. Wavelength** corresponded of maximum energy density (Wien's Law) is
$$\lambda_0 = \frac{b}{T}, \quad \omega = \frac{2\pi}{\lambda_0}, \qquad (3)$$
where $b = 2.8978 \times 10^{-3}$ is constant, m °K; $T$ is temperature, °K; $\omega$ is angle frequency of wave, rad/s.

**5. Pressure for single full reflection** is
$$F = 2E/c, \qquad (4)$$

where $F$ - pressure, N/m$^2$; $c = 3\times10^8$ is light speed, m/s, $E$ is radiation power, W/m$^2$. If plasma does not reflect radiation the pressure equals

$$F = E/c. \tag{5}$$

**6. Pressure for plasma multi-reflection** [23-25] is

$$F = \frac{2E}{c}\left(\frac{2}{1-q}\right), \tag{6}$$

where $q$ is plasma reflection coefficient. For example, if $q = 0.98$ the radiation pressure increases by 100 times. We neglect losses of prism reflection.

**7. The Bremsstrahlung (brake) loss** energy of plasma by radiation is ($T > 10^6$ °K)

$$P_{Br} = 5.34 \cdot 10^{-37} n_e^2 T^{0.5} Z_{eff}, \quad \text{where} \quad Z_{eff} = \sum (Z^2 n_z)/n_e, \tag{7}$$

where $P_{Br}$ is power of Bremsstrahlung radiation, W/m$^3$; $n_e$ is number of particles in m$^3$; $T$ is a plasma temperature, KeV; $Z$ is charge state; $Z_{eff}$ is cross-section coefficient for multi-charges ions. For reactions H+D, D+T the $Z_{eff}$ equals 1.

That loss may be very much. For some reaction they are more then useful nuclear energy and fusion nuclear reaction may be stopped. The Bremsstrahlung emission has continuous spectra.

**8. Electron frequency in plasma** is

$$\omega_{pe} = \left(\frac{4\pi n_e e^2}{m_e}\right)^{1/4}, \quad \text{or} \quad \omega_{pe} = 5.64\times10^4 (n_e)^{1/4}, \tag{8}$$

in "cgs" units, or $\omega_{pe} = 56.4(n)^{1/4}$ in CI units

where $\omega_{pe}$ is electron frequency, rad/s; $n_e$ is electron density, [1/cm$^3$]; $n$ is electron density, [1/m$^3$]; $m_e = 9.11\times10^{-28}$ is mass of electron, g; $e = 1.6\times10^{-19}$ is electron charge, C.

The plasma is reflected an electromagnet radiation if frequency of electromagnet radiation is less then electron frequency in plasma, $\omega < \omega_{pe}$. That reflectivity is high. For $T > 15\times10^6$ °K it is more than silver and increases with plasma temperature as $T^{3/2}$. The frequency of laser beam and Bremsstrahlung emission are less then electron frequency in plasma.

**9. The deep of penetration** of outer radiation into plasma is

$$d_p = \frac{c}{\omega_{pe}} = 5.31\cdot10^5 n_e^{-1/2} \text{ [cm]} \tag{9}$$

For plasma density $n_e = 10^{22}$ 1/cm$^3$ $d_p = 5.31\times10^{-6}$ cm.

**10. The gas (plasma) dynamic pressure**, $p_k$, is

$$p_k = nk(T_e + T_i) \quad \text{if} \quad T_e = T_k \quad \text{then} \quad p_k = 2nkT, \tag{10}$$

where $k = 1.38\times10^{-23}$ is Boltzmann constant; $T_e$ is temperature of electrons, °K; $T_i$ is temperature of ions, °K. These temperatures may be different; $n$ is plasma density, 1/m$^3$; $p_k$ is plasma pressure, N/m$^2$.

**11. The gas (plasma) ion pressure**, $p$, is

$$p = \frac{2}{3}nkT, \tag{11}$$

Here $n$ is plasma density in 1/m$^3$.

**12. The magnetic $p_m$ and electrostatic pressure**, $p_s$, are

$$p_m = \frac{B^2}{2\mu_0}, \quad p_s = \frac{1}{2}\varepsilon_0 E_s^2, \tag{12}$$

where $B$ is electromagnetic induction, Tesla; $\mu_0 = 4\pi \times 10^{-7}$ electromagnetic constant; $\varepsilon_0 = 8.85 \times 10^{-12}$, F/m, is electrostatic constant; $E_S$ is electrostatic intensity, V/m.

**13. Ion thermal velocity** is

$$v_{Ti} = \left(\frac{kT_i}{m_i}\right)^{1/2} = 9.79 \times 10^5 \mu^{-1/2} T_i^{1/2} \quad \text{cm/s}, \quad (13)$$

where $\mu = m_i/m_p$, $m_i$ is mass of ion, kg; $m_p = 1.67 \times 10^{-27}$ is mass of proton, kg.

**14. Transverse Spitzer plasma resistivety**

$$\eta_\perp = 1.03 \times 10^{-2} Z \ln \Lambda T^{-3/2}, \quad \Omega \text{ cm} \quad \text{or} \quad \rho \approx \frac{0.1 Z}{T^{3/2}} \quad \Omega \text{ cm}, \quad (14)$$

where $\ln \Lambda = 5 \div 15 \approx 10$ is Coulomb logarithm, $Z$ is charge state.

**15. Reaction rates** $<\sigma v>$ (in cm$^3$ s$^{-1}$) averaged over Mexwellian distributions for low energy (T<25 keV)

may be represent by

$$\begin{aligned}(\overline{\sigma v})_{DD} &= 2.33 \times 10^{-14} T^{-2/3} \exp(-18.76 T^{-1/3}) \quad \text{cm}^3\text{s}^{-1}, \\ (\overline{\sigma v})_{DT} &= 3.68 \times 10^{-12} T^{-2/3} \exp(-19.94 T^{-1/3}) \quad \text{cm}^3\text{s}^{-1},\end{aligned} \quad (15)$$

where T is measured in keV.

**16. The power density** released in the form of charged particles is

$$\begin{aligned}P_{DD} &= 3.3 \times 10^{-13} n_D^2 (\overline{\sigma v})_{DD}, \quad \text{W cm}^{-3} \\ P_{DT} &= 5.6 \times 10^{-13} n_D n_T (\overline{\sigma v})_{DT}, \quad \text{W cm}^{-3} \\ P_{DHe^3} &= 2.9 \times 10^{-12} n_D n_{He^3} (\overline{\sigma v})_{DHe^3}, \quad \text{W cm}^{-3}\end{aligned} \quad (16)$$

Here in $P_{DD}$ equation it is included D + T reaction.

## Computed estimations of AB-Reactors

We consider two new Micro-AB-Reactors having the innovative: multi-reflect radiation and self-magnetic confinement features.

In multi-reflect radiation confinement of AB-Reactor the offered innovation is the special prisms, a high reflectivity mirror that returns the laser beam exactly to its point of origination. As a result, the all energy absorbs by plasma, the laser radiation multi-times presses the plasma and impedes it or, at least, it does not allow its expansion. The plasma has high reflectivity and this press effect may be increased hundreds to thousands of times. Practically speaking, we are weakly limited and can use the cheap and solid fuel.

The uniformly heating of target by laser beam is very big and important problem in ICF. The non-equal rocket forces on target shell destroy the capsule before thermonuclear ignition. If the first ICF reactor had some laser beams, the second generation had 10 laser beams (NOVA), the third generation has 60 beams (OMEGA), and the next generation will have 192 beams (NIF). All laser beams must be equal and work in coordination - that is complex, difficult and expensive problem. The prism reflector is easy designed such the reflected beam runs round the target and presses it uniformly from all sides after 2 - 4 reflections.

The second innovation is the special form microcapsule that is filled by compression gas or liquid (frozen) fuel.

In self-magnetic confinement Micro-AB-Reactor main innovation is super thin microcapsule and electric heating which produce high-intensity magnetic field, keeping the plasma pressure and conic (or close to conic) ends of ampoule capsule which work as plasma stoppers. The important innovation is the using an electric currency for straight heating of capsule. The magnetic lines in our reactor are circles located into and around plasma channel. The magnetic intensity increases from central axis to maximal plasma radius. That pushes plasma into center of plasma channel. In the ends of plasma channel the magnetic forces put obstacles in plasma leakage.

For estimation possibilities of these innovations in the first AB-reactor we compute the multi-reflection pressure, the condition of plasma reflection and compare them with dynamic pressure of plasma. In the second AB-reactor we consider the equilibrium the magnetic and kinetic pressures.

**Capsules**. For multi-reflex conformation is more suitable the *spherical capsule*. Let us consider the gas compressed fuel capsule. The shell thickness and relative weight of gas compressed fuel spherical capsule can be computed by equations:

$$\frac{\delta}{r} = \frac{P}{2\sigma}, \quad M_R = \frac{3}{2}\frac{P}{\sigma}\frac{\gamma_s}{\gamma_f}, \quad (17)$$

where $\delta$ is shell thickness of capsula, [m]; $r$ is capsule radius, [m[; $\delta/r$ is relative thickness of fuel shell; $P$ is fuel gas pressure into capsule (over the atmospheric pressure), [N/m$^2$]; $\sigma$ is safety tensile stress, N/m$^2$; $M_R$ is relative capsule mass; $\gamma_s$ is density of capsule shell, [kg/m$^3$]; $\gamma_f$ is density of fuel, [kg/m$^3$].

Example, for gas pressure $P = 100$ atm $= 10^7$ N/m$^2$, $\sigma = 200$ kgf/mm$^2 = 2\times10^9$ N/m$^2$, $\gamma_s = 1800$ kg/m$^3$; $\gamma_f = 11$ kg/m$^3$ (at $P = 100$ atm) we get $\delta/r = 2.5\times10^{-3}$, $M_R = 0.6$.

*Cylindrical capsula* ($l \gg r$). For our estimations we take the capsule having the length 1 mm, radius $r = 0.05$ mm, cross section area $S \approx 8\times10^{-3}$ mm$^3$, fuel volume $V_c = 8\times10^{-3}$ mm$^3 = 8\times10^{-12}$ m$^3$. That is very small. It is a microcapsule.

If the gas in a microcapsule is pressed the relative thickness and relative mass of cylindrical shell may be computed by equations:

$$\frac{\delta}{r} \approx \frac{P}{\sigma}, \quad M_R \approx 2\frac{P}{\sigma}\frac{\gamma_s}{\gamma_f}, \quad (18)$$

For $P = 100$ atm ($P = 10^7$ N/m$^2$) and $\sigma = 200$ kg/mm$^2$ ($\sigma = 2\times10^9$ N/m$^2$) ratio $\delta/r = 5\times10^{-3}$, $M_S = 1.3$.

**Fuel density**. The particles (ions) density $n$ of fuel in 1 m$^3$ and number of particles $n_C$ in microcapsule equal:

$$n = \frac{1}{2m_p}\left(\frac{\gamma_{f1}}{\mu_1} + \frac{\gamma_{f2}}{\mu_2}\right) = \frac{\gamma_{fa}}{\mu_a m_p} = \frac{\gamma_{fa}}{m_{ia}} \approx \frac{\gamma_{f1}}{\mu_1 m_p}, \quad n_c = nV_c$$

$$\text{where} \quad \mu_1 = \frac{m_{i1}}{m_p}, \quad \mu_2 = \frac{m_{i2}}{m_p} \quad (19)$$

where $m_{ia} = 2.5\times1.672\times10^{-27}$ kg for D+T is average mass of fuel ion; $m_p = 1.672\times10^{-27}$ kg is proton mass; low indexes "$_{1, 2}$" means the first and the second fuel component.

The $n \approx 10^{20}$ 1/m$^3$ in the present magnetic confinement fusion reactor; $n \approx 2.6\times10^{25}$ 1/m$^3$ for gas D+T in a pressure 1 atm, $T = 288$ °K (the density of deuterium D is $\gamma_{f1} = 0.0875$ kg/m$^3$, $\mu_1 = 2$, the density of Tritium is 1.5 more). For D and other pressure the $n$ must be changed in same times, for example, if $P = 100$ atm then $\gamma_f = 8.75$ kg/m$^3$, $n \approx 2.6\times10^{27}$ 1/m$^3$, $n_c = 20.8\times10^{15}$; $n \approx 2.1\times10^{28}$ 1/m$^3$, $n_c = 1.7\times10^{17}$ ($\gamma_f = 71$ kg/m$^3$) for liquid hydrogen at a pressure of 1 atm. (In conventional inertial confinement fusion reactor, the fuel density may be more in 10 - 30 times, under a rocket pressure of a fuel capsule cover. For hydrogen the frizzed temperature is -259.34 °C, the boiling point is -252.87 °C). For D+T average $\mu_a = (2+3)/2 = 2.5$.

**Fuel mass $M_f$ [kg] and thermonuclear energy $E_C$ into microcapsule** are computed by equations:

$$E = 1.6 \cdot 10^{-19} \frac{M_f}{\mu_N m_p} E_r = 0.96 \cdot 10^8 M_f \frac{E_r}{\mu_N},$$

$$M_f = \mu_N m_p n V_C, \quad E_C = 0.5 \times 1.6 \cdot 10^{-19} n V_C E_R \quad (20)$$

where $E_r$ is reaction energy of one couple particles, eV; $\mu_N = \mu_1 + \mu_2$ is number of nucleons which take part in reaction, for D+T $\mu_N = 2+3=5$. If we want compute energy of one type of particles, the $E_r$ is reaction energy for given type of particles, for D+T the energy $E_r = 17.5$ MeV;. For example, our capsule (1×0.1 mm) filled by liquid fuel D+T has fuel mass $M_f = 0.71 \times 10^{-3}$ μg and will produce energy $E_c = 240$ kJ. If we burn out 10 capsules per 1 s, the engine power will be 2400 kW. An estimated 20% this energy gives the charged particles and 80% of neutrons. The fuel capsule having $M = 10$ μg $=10^{-5}$ kg of a mixture D+T produces $3.34 \times 10^9$ J if all atoms take part in reaction. That is equivalent to the energy derived from 84 liters of benzene.
Computations are presented in fig. 9.

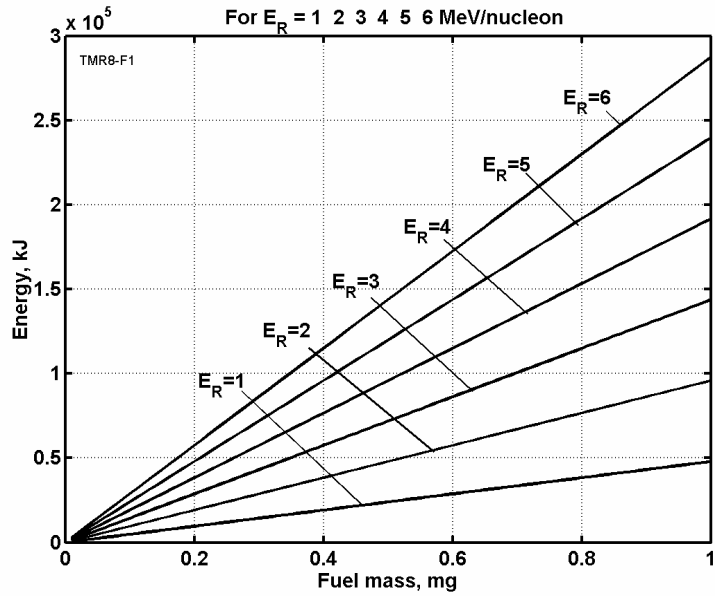

**Fig. 9.** Energy of thermonuclear reactor versus the fuel mass and energy per one nucleon. $E_R = E_r/\mu_N$.

**Distribution of thermonuclear energy between particles**. In most cases the result of thermonuclear reaction is two components. As you see in Table 1 that may be "He" and neutron or proton. The thermonuclear energy distributes between them the following manner:

$$\text{From} \quad E = E_1 + E_2 = \frac{m_1 V_1^2}{2} + \frac{m_2 V_2^2}{2}, \quad m_1 V_1 = m_2 V_2,$$

$$\text{we have} \quad \frac{E_1}{E} = \frac{m_2}{m_1 + m_2} = \frac{\mu_2}{\mu_1 + \mu_2}, \quad E_2 = E - E_1 \quad (21)$$

where $m$ is particle mass, kg; $V$ is particle speed, m/s; $E$ is particle energy, J; $\mu = m_i/m_p$ is relative particle mass. Lower indexes "$1, 2$" are number of particles.

Unfortunately, as you can see (in Table 1), most particle energy catches the neutron as the lightest particle. But its emission has high penetrating capability, creating radioactive isotopes, causing damage to the main construction, very dangerous for living beings, and that can be converted only in heat energy.

**Energy is needed for fuel heating**. This energy can be estimated by equation:

$$E_h = \frac{c}{\mu_a} MT_k, \quad c = \frac{k}{2m_p} = 4.13 \cdot 10^3, \quad \mu_a = \frac{m_{ia}}{m_p}, \quad (22)$$

where $c$ is plasma thermal capacity, J/kg·°K; $T_k$ is temperature in °K; $k$ is Boltzmann constant, $m_{ia}$ is average mass of ions, kg; $M$ is fuel mass, kg. This computation is presented in fig. 10. Our capsule filled by liquid mixture D+T requests ignition energy about 124 J for its heating up to 100 million °K. That is energy of electric condenser having size about 10×10×10 cm for $\beta = 10^8$ V/m (see below).

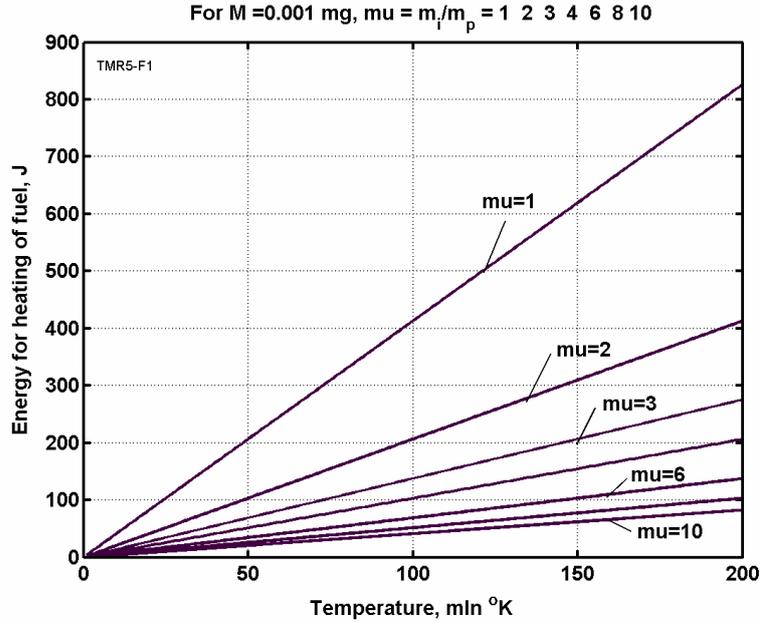

**Fig. 10.** Energy requested for fuel D+T heating.

**Capacitor for thermonuclear ignition.** Condenser requested as storage of energy for fuel thermonuclear ignition my be estimated by equation

$$\frac{W}{V} = \frac{1}{2}\varepsilon_0 \varepsilon \beta^2, \quad \frac{W}{M} = \frac{\varepsilon_0 \varepsilon \beta^2}{2\gamma}, \quad (23)$$

where $W$ is condenser energy, J; $V$ is condenser volume, m³; $M$ is condenser mass, kg; $\varepsilon_o = 8.85 \times 10^{-12}$ Φ/m is the electrostatic constant; $\varepsilon$ is dielectric constant; $\beta$ is dielectric strength, V/m; $\gamma$ is specific density of dielectric, kg/m³. Energy from capacitor is about one joule from one centimeter cub.

**Electron plasma frequency**. Electron frequency of plasma is computed by equation (8). For $n \approx 10^{20}$ 1/m³ that is equals $\omega_{pe} = 5.64 \times 10^{11}$ rad/s, for $n \approx 10^{28}$ 1/m³ that equals $\omega_{pe} = 5.64 \times 10^{15}$ rad/s. That is more then the laser frequency ($\lambda = 0.3 \times 10^{-9}$ m, $\omega = 2.1 \times 10^{10}$ rad/s). That means the plasma will reflect the laser beam.

**Plasma skin depth**. The depth in plasma to which an electromagnetic radiation can penetrate (Eq. (9)) is: For $n \approx 10^{20}$ 1/m³ that is equals $d_p = 5.31 \times 10^{-2}$ cm, for $n \approx 10^{28}$ 1/m³ that equals $d_p = 5.31 \times 10^{-6}$ cm. As you see, the depth is small.

**Coefficient reflectivity of plasma**. No data about plasma reflectivity. However, from general theory of reflectivity it is known the reflectivity depends from conductivity (see Eq. (14)). Silver has the best conductivity from solid body and best reflectivity. It is about $q = 0.78 \div 0.99$ (it depends from frequency of radiation: for ultra-violet radiation $q \approx 0.78$, for thermal radiation $q \approx 0.99$). The plasma for $T > 15 \times 10^6$ °K has better conductivity then silver. The plasma conductivity increases as $T^{3/2}$. That means the plasma having the $T \approx 10^8$ °K has reflectivity in 17.2 times better then silver. That means the plasma reflectivity is more 0.999. We take in our computation $q = 0.999$. The efficiency of offered innovation very strong depends from reflectivity of plasma. The reflectivity of the prism mirror is very high [23]. We neglect the loss in it.

**Bremsstrahlung (brake) radiation.** That is proportional the energy spectra $E$ and has Maxwell distribution:

$$f_E dE = f_p \left(\frac{dP}{dE}\right) dE = 2\sqrt{\frac{E}{\pi(kE)^3}} \exp\left[\frac{-E}{kT_k}\right] dE,$$

$$\lambda = \frac{c}{\nu}, \quad \nu = \frac{c}{\lambda} = \frac{E}{h}$$

(24)

where $k = 1.38 \times 10^{-23}$ is Boltzmann's constant, J/°K; $T$ - temperature, °K; $E$ - energy, J; $\nu$ - frequency, 1/s; $\lambda$ length of wave, m; $h = 6.525 \times 10^{-34}$ is Planck's constant, J's. Assume the brake radiation has same specter.

Computations are presented in figs. 11-12.

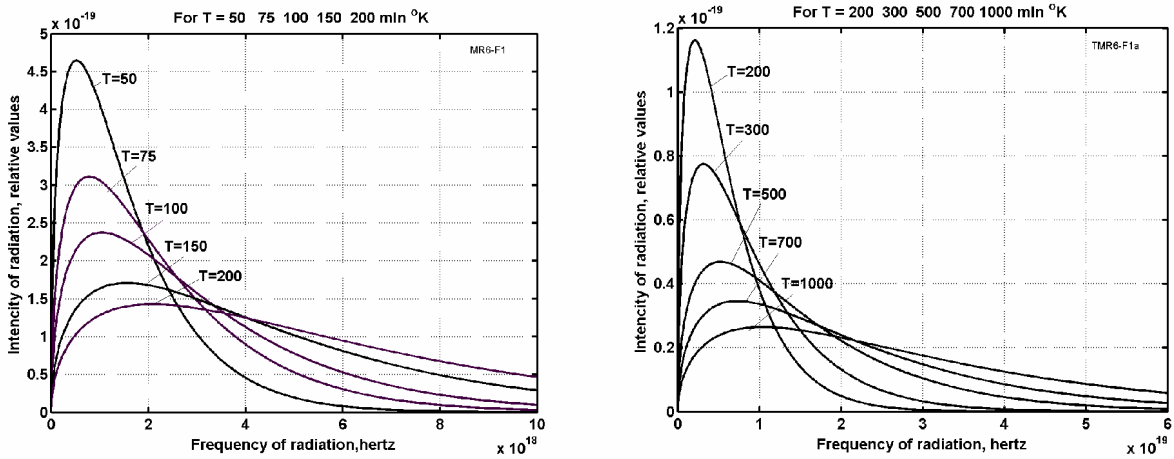

**Fig. 11** (left). Spectra of brake radiation for plasma temperature 60 - 200 millions degrees (°K).
**Fig. 12** (right). Spectra of brake radiation for plasma temperature 200 - 1000 millions degrees (°K).

The ultra-violet rays are below approximately $< 3 \times 10^{17}$ hertz ($\lambda > 10^{-9}$ m), the soft x-rays are below $< 3 \times 10^{19}$ hertz ($\lambda > 10^{-11}$ m). That means the brake radiation can be reflected by special methods. For example, the silver having high electro-conductivity has average reflectivity 0.99 in heat region, 0.95 in light region, and 0.78 in ultra-violet region. Some metals has reflectivity up 0.2 for $\lambda = 40 \times 10^{-9}$ m. But plasma having the temperature more than $15 \times 10^6$ °K has more electro-conductivity then silver and it must, therefore, have better reflectivity. The reflectivity coefficient of prism mirror is very high and we can neglect its losses. However, the reflectivity of prism mirror for brake radiation is needed in a detailed test.

The average energy of Bremsstrahlung photon equals average energy of plasma electron. The formula for average wavelength is:

$$\text{From} \quad E = kT_k = h\nu, \quad \lambda = \frac{c}{\nu}$$

$$\text{we receive} \quad \lambda_e = \frac{ch}{kT_k} = \frac{0.0144}{T_k}$$

(25)

where $E$ is electron energy, $c = 3 \times 10^8$ is light speed, m/s; $T_k$ is temperature in °K; $\lambda_e$ is wave length, m ; $\nu$ is wave frequency, 1/s;.

For example, for $T_k = 10^8$ °K the $\lambda_e = 1.44 \times 10^{-10}$ m. That value is the lower ultra-violet diapason $\lambda > 10^{-9}$ m.

For very high temperature the most part of this spectrum is in the soft x-ray region, but soft x-ray can be reflected and retracted by special methods.

**The reactive pressure**. We can estimate that the ion speed for $T = 10^8$ °K. That is approximately $V = 600$ km/s. If $M = 0.1$ $\mu g = 10^{-7}$ kg of a mixture D+T is increased their speed to this value in time $\tau = 10^{-9}$ s, the reactive force will be $F = MV/\tau \approx 5 \times 10^7$ N. If the fuel capsule has surface $s = 5$ mm$^2 = 5 \times 10^{-6}$ m$^2$ the capsule cover pressure is $p = F/s = 10^{13}$ N/m$^2 = 10^8$ atm. This pressure produces the shockwave which compresses the fuel microcapsule and create high ion temperature.

## Radiation confinement

Radiation confinement is suitable for multi-reflex laser beam support.
**Equilibrum of Multi-reflex laser and kinetic pressures.**
. From equations (6), (10) we receive

$$P_k = 2nkT_k, \quad P_R = \frac{2E}{c}\left(\frac{2}{1-q}\right), \quad P_k = P_R,$$

$$P_L = SE = 0.5nkT_k cS(1-q) \tag{26}$$

where $P_L$ is impulse power of laser, W; $S$ is surface of capsule or plasma; $q$ is plasma reflection. The additional number 2 appears because we neglect the prism reflection loss. The computations for $n = (0 \div 1) \times 10^{28}$ 1/m$^3$, $S = (1 \div 4) \times 10^{-6}$ m$^2$, $q = 0.999$, $T_k = 10^8$ are presented in Fig. 13.

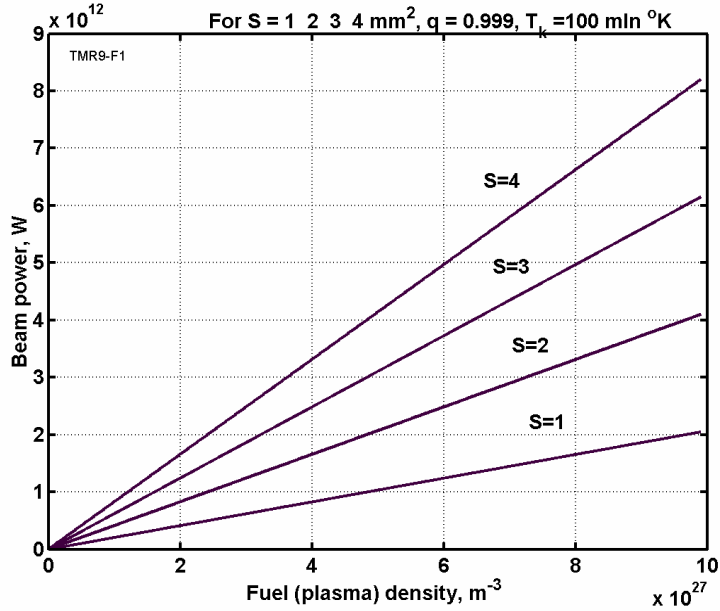

**Fig**. 13. Equilibrium of multi-reflex laser pulse power versus plasma density and fuel capsule surface $S$ for coefficient plasma reflectivity $q = 0.999$, plasma temperature $T_k = 10^8$ °K.

Look your attention that power of laser pulse for multi-reflection confinement is in tens - hundreds times less then one is in the current ICF reactors (OMEGA has $60 \times 10^{12}$ W, Z-machine will have $350 \times 10^{12}$ W). That shows, the multi-reflect conformation is more efficiency then rocket conformation for small targets.

We can increase the initial multi-reflex pressure in millions times if we cover the outer capsule surface the small reflective prism as internal surface of the combustion chamber. As it is shown in [26]

p. 378, fig. A3.4, the pressere from 1 kW of laser power can reach more $10^4$ N. If laser pulse power has $P_L = 10^{13}$ W, the pressure will be $F = 10^{17}$ N. The surface of a spherical capsule having diameter 1 mm is about $S = 3$ mm$^2 = 3 \times 10^{-6}$ m$^2$. Hence the pressure on target is $P = F/S = 3.3 \times 10^{22}$ N/m$^2 = 3.3 \times 10^{17}$ atm! That is in $10^9$ times more then a gas dynamic pressure of the plasma at temperature $10^8$ K.

  **Note.** The rocket force used for compressing and heating pullet at present time cannot keep the big pressure
and temperature for very small capsules at need time because gas extension. For example, let us to estimate the extension time for two capsules having diameter $d = 0.3$ mm and $d = 3$ mm respectively at temperature $10^8$ K. The average ion speed at this temperature is about $6 \times 10^5$ m/s. For typical pulse time $10^{-9}$ s the plasma radius is increased in $6 \times 10^{-4}$ m = 0.6 mm. That means the volume of the first capsule increases in $(0.75/0.15)^3 = 125$ times, the volume of the second capsule is increased only in $(2.1/1.5)^3 = 2.7$ times. In our multi-reflex reactor the beam pressure does not allow to expansion the plasma and increases the reaction time and possibility thermonuclear reaction in hundreds times.

**Requested minimum time of laser pulse.** Duration of laser pulse needed for heating of capsule can be computed by equation

$$\tau = \frac{c_1 M_f T_k}{P_L}, \qquad (27)$$

where $c_1 = 4.13 \times 10^3$ is thermal capacity coefficient, J/kg.K; $T_k$ is plasma temperature, K. The computations are presented in Fig. 14.

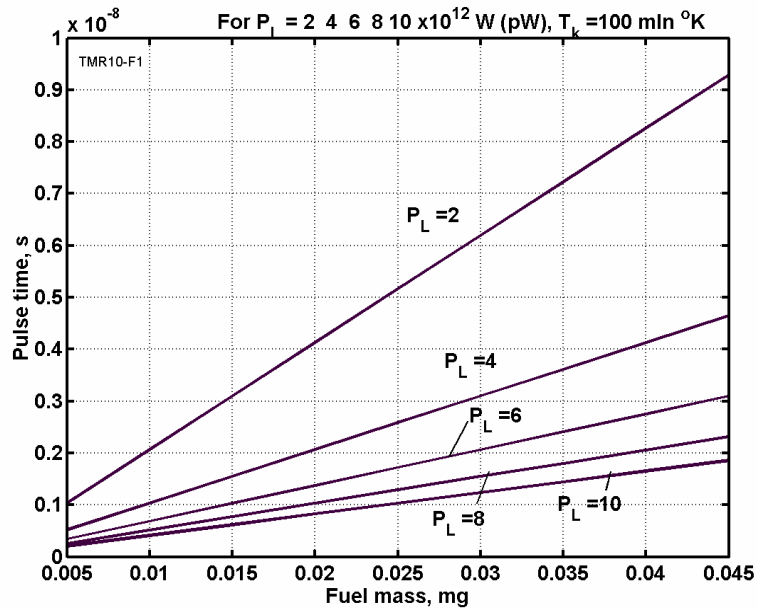

Fig. 14. Request laser pulse time for capsule heating via capsule mass and laser pulse power for capsule temperature $10^8$ K.

As you see, the pulse is same with current laser $(1 \div 10) \times 10^{-12}$ s, (ps). In conventional ICF reactor the most part beam energy is reflected by plasma and heating the shell of combustion chamber. In our reactor the nearly all beam energy is used for capsule heating.

**Equilibrium of brake radiation and kinetic pressures.**
From equations (10), (7) we receive

$$P_k = 2nkT_k, \quad P_{BP} = \frac{2V}{c(1-q)S} 5.34 \cdot 10^{-37} n^2 T_e^{1/2} Z_{eff}, \quad P_k = P_{BP},$$

$$n = \frac{k\,c(1-q)S}{5.34 \cdot 10^{-37} V Z_{eff}} \frac{T_k}{T_e^{1/2}} = 9 \cdot 10^{28}(1-q)T_e^{1/2} \frac{S}{V Z_{eff}}$$

(28)

where $T_e$ is temperature in keV; $V$ is plasma volume, m³; $S$ is plasma surface, m²; q is average coefficient reflectivity of x-rays produced by brake radiation. The equilibrium of brake radiation and kinetic pressure can be reached for ratio $V/S \approx 1$.

## The self-magnetic confinement.

The self-magnitude confinement is suitable for low-density plasma. Your attention is called for to the big difference between a present conventional reactor magnetic field and the offered self-magnetic field. For creating of the present magnetic field, the large powerful superconductivity very expensive magnets are used. Our self-magnetic does not request anything. The self-magnetic field is produced by capsule electric current and that is more powerful by hundreds of times. Why? The magnetic intensity and pressure of electric current in inverse proportion of plasma radius (see equations (29) below). The present thermonuclear reactor has plasma camera of some meters. Our capsule has radius only 0.05 mm.

**Equilibrium of self-magnetic and kinetic plasma pressure**.
From equation (10) and (12) we receive

$$P_k = P_m, \quad P_k = 2nkT_k, \quad P_m = \frac{\mu_0 H^2}{2}, \quad H = \frac{I}{2\pi r}, \quad P_m = \frac{\mu_0}{8\pi^2}\left(\frac{I}{r}\right)^2,$$

$$I = 4\pi r \sqrt{\frac{knT_k}{\mu_0}} = 4.16 \cdot 10^{-8} r (nT_k)^{0.5}, \quad U = RI, \quad B = \frac{\mu_0 I}{2\pi r}$$

(29)

where $r$ is radius of capsule (plasma flux), m; $I$ is electric currency, A; $R$ is capsule (plasma) resistance, Ohm; $U$ is capsule (electrodes) voltage, V; $H$ is magnetic intensity, A/m; $B$ is magnetic intensity, Tesla; $T_k$ is plasma temperature, °K.

The computations for several $n$ are presented in figs. 15 - 18.

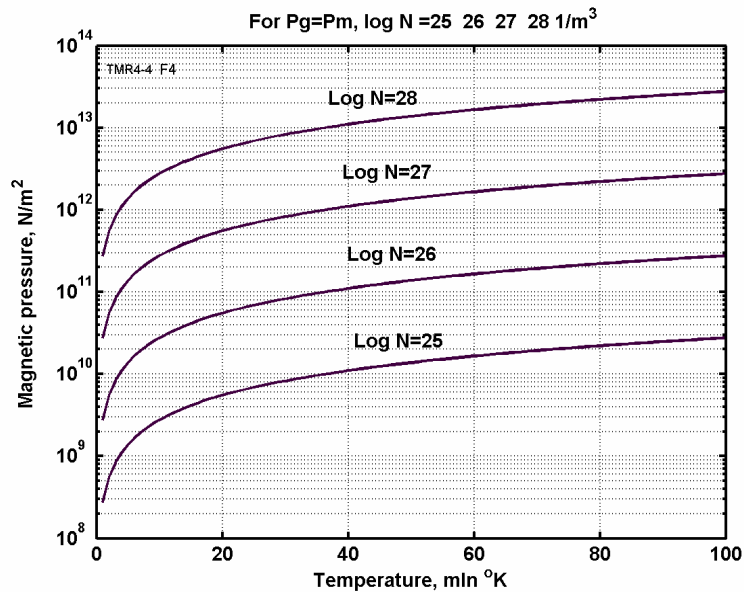

**Fig. 15.** Equilibrium self-magnetic and kinetic pressures versus plasma temperature and plasma densities. Capsule 0.1×1 mm. *N* is plasma density, 1/m$^3$.
**Note:** This pressure is same for multi-reflex and plasma pressure.

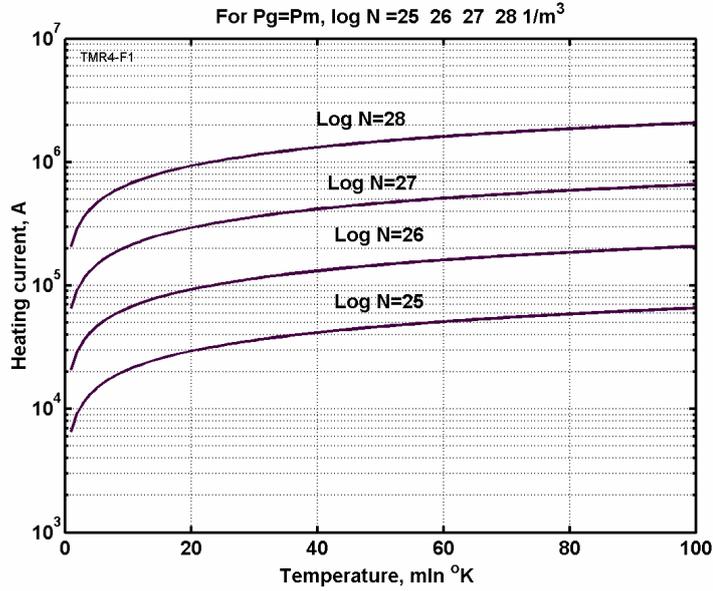

**Fig. 16.** Electric currency needed for equilibrium kinetic and magnetic pressure for several plasma densities.
Microcapsule has size 0.1×1 mm. *N* is plasma density, 1/m$^3$.

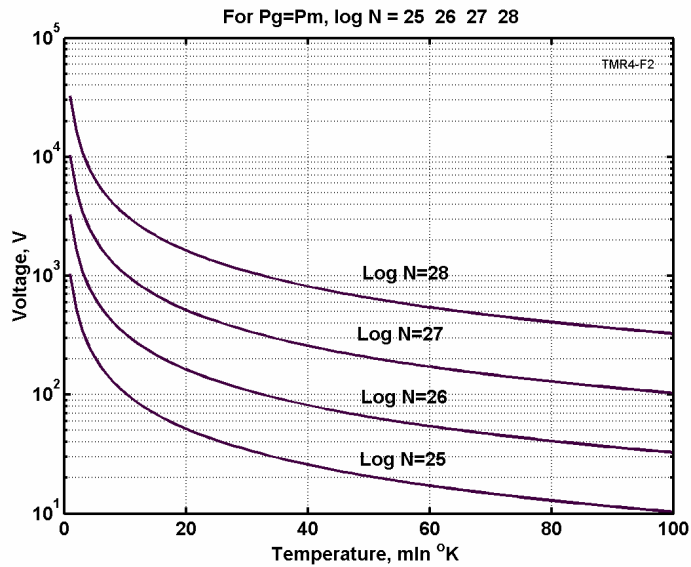

**Fig. 17.** Electric voltage needed for equilibrium kinetic and magnetic pressures for several plasma densities.
Capsule has size 0.1×1 mm. *N* is plasma density, 1/m$^3$.

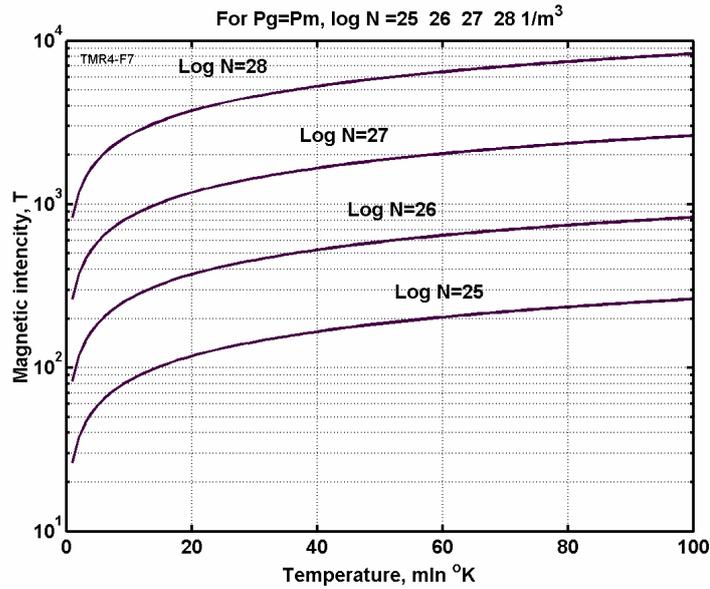

**Fig. 18.** Equilibrium self-magnetic intensity on microcapsule surface via plasma temperature for several plasma densities. Capsule has size 0.1×1 mm. $N$ is plasma density, $1/m^3$.

The present magnetic confinement reactor having superconductivity magnets has maximum magnetic intensity 5 - 6 Tesla. As you see in fig. 18 the offered AB-self-magnetic reactor has more magnetic intensity in hundreds times.

## Projects

Below there are estimations of some projects, which show parameters of suggested AB-reactors. These are not optimal reactors. They demonstrate the methods of computations and possible technical data of new micro reactors.

**1. Multi-reflection AB-reactor.** Let us to take the spherical fuel capsule diameter $d$ = 1 mm. Its surface is 3.14 mm$^2$, the volume is $v$ = 0.52 mm$^3$=5,2×10$^{-10}$ m$^3$. If gaseous fuel (D+T) has pressure 1, 10, 100 atm, the specific fuel density are $\rho$ = 0.11 kg/m$^3$, 1.1 kg/m$^3$, 11 kg/m$^3$ respectively. The fuel mass are $M_f = \rho v$ = 5.7×10$^{-11}$, 57×10$^{-11}$, 570×10$^{-11}$ kg respectively. Particle densities are $n_1 = \rho/\mu_a m_p$ = 2.63×10$^{25}$ 1/m$^3$, 2.63×10$^{26}$ 1/m$^3$, 2.63×10$^{27}$ 1/m$^3$ respectively. Numbers of particles in the capsule are $n = n_1 v$ = 1.37×10$^{16}$, 1.37×10$^{17}$, 1.37×10$^{18}$ respectively.

Thermonuclear energy in capsule are $E = 0.5 n E_1$ = 1.9×10$^4$ J, 1.9×10$^5$ J, 1.9×10$^6$ J respectively. Here $E_1$ = 17.6×10$^6$×1.6×10$^{-19}$ = 2.8×10$^{-12}$ J is the energy in single reaction of couple particles. Where 17.6×10$^6$ MeV is thermonuclear energy of reaction D+T. If we burn 1 capsule in 1 second, the thermonuclear power will be $W$ = 1.9×10$^4$ W, 1.9×10$^5$ W, 1.9×10$^6$ W respectively.

Fuel heating energy are $E_f = c_1 M_f T_k$ = 24 J, 240 J, 2400 J respectively. Here $c_1$ = 4.13×10$^3$ is average thermal capacity of plasma, $T_k$ = 10$^8$ K is maximal plasma temperature. These heating energy must be increased some (3 ÷ 6) times because we must to heat the capsule shell and coefficient of energy efficiency is less then 1. The current condensers have energy storage capability about 1 J/cm$^3$.

Requested minimum (equilibrium) pulse laser power equal $N$ = 17.1×10$^9$ W, 17.1×10$^{10}$ W, 17.1×10$^{11}$ W respectively (Eq. (22)) for $q$ = 0.999. Pulse time is $\tau = E_f/N$ = 1.4×10$^{-9}$ s.

We can use the liquid fuel. All parameter significantly will improvement (approximately in 10 times with comparison of the 100 atm capsule), but we get a problem with storage of capsules into a liquid helium.

**2. Self-magnetic AB-reactor**. Let us to take the fuel capsule of the length $L = 1$ mm, diameter $2r = 0.1$ mm and gaseous fuel (D+T) pressure $p = 100$ atm . The cross-section of capsule is $S = 7.85 \times 10^{-3}$ mm$^2$, volume $v = 7.85 \times 10^{-12}$ m$^3$, fuel mass is $M_f = \rho v = 9.5 \times 10^{-11}$ kg, particle density is $n_1 = \rho / \mu_a \cdot m_p = 2.63 \times 10^{27}$ 1/m$^3$, number of particle into capsule is $n = n_1 v = 2.06 \times 10^{16}$. Heating fuel energy is $E_f = c_1 M_f T_k = 39$ J, for $T_k = 10^8$ K. If we burn 1 capsule in 1 second the thermonuclear power will be $W = 3 \times 10^4$ W.

Requested minimum (equilibrium) electric currency equal $I = 1.07 \times 10^6$ A (for $T_k = 10^8$ K),(Eq. (29)). The plasma electric specific resistance at $T_k = 10^8$ K is $\rho = 0.1 Z/T^{3/2} = 1.23 \times 10^{-7}$ Ω sm. The electric plasma resistance is $R_f = \rho L/S = 1.5 \times 10^{-4}$ Ω (all for $T_k = 10^8$ K). Voltage $U = IR_f = 160$ V. Pulse power $N = IU = 17.1 \times 10^7$ W. Pulse time is $\tau = 2.3 \times 10^{-6}$ s. Maximum intensity of magnetic field is $B = \mu_o I / 2\pi r = 4280$ T.

## Discussion

The offered thermonuclear AB-Reactors, as with any innovations, are needed in further more detailed laboratory research, product development and testing. However, theses new Reactors have gigantic advantages over present-day thermonuclear reactors:

1. They are cheaper by many hundreds of times. That means not only non-industrial countries but middle-size companies can undertake R&D and production of perfected new thermonuclear reactors.
2. They have a small weight and size but they have enough power (up 10,000 kW). That means they can be used as engine of land vehicles, small ships, aircraft, manned and unmanned spacecraft propulsion and community power utilities.
3. They are not limited in high temperature regime as are all existing reactors. That means they can use inexpensive fuel (not deuterium, tritium, helium-3, uranium as do extant reactors).

The parameters of AB-Reactors are considered and computed in given article very far from optima. They are only examples utilized to vividly illustrate the large possibilities of the innovative reactors.

The suggested AB-thermonuclear reactor has Lawson criterion in some order more then conventional current (2005) thermonuclear reactors (ICF). That strongly increases either of three multipliers in Lawson criterion. That increases the density $n$ up to two-three orders. It increases the temperature $T$. It returns the laser and thermal radiation back to fuel pellet. (This emission is lost in present reactors). It increases the time of reaction $\tau$. The suggested AB thermonuclear reactors may be a revolutionary jump in energy industry.

**Note**: In conventional ICF the initial (internal into plasma) radiation does not compress the plasma. Plasma is transparency for internal radiation. That emission influences only to an emitted particle. When radiation came out of source (fuel pellet) and reflected or adsorbed by chamber surface that does not press on pellet surface. By that means, the conventional inertial thermonuclear reactor has only losses from radiation. The offered AB Reactor has the big desirable benefits from thermal radiation. The more radiation, the more benefits.

The offered AB-Reactor can also have problems. The radiation mirror can have a bad reflectivity for ultra-violent rays or experimenters may have problems with fast high-intensity electric impulse through small capsule. However, if mirror will be reflect only conventional ultra-violet, light, and thermal radiations that would be enough for ignition of a thermonuclear reaction. As any innovation the offered reactor needs further perfecting R&D.

The offered AB-self magnetic reactor is different from present magnetic confinement reactor. That is smaller because AB-self-magnetic reactor works a small fuel capsule. In present-day reactor, the rare fuel gas (D+T) fills all volume of large chamber. In AB-Reactor the fuel is located into small capsule under high pressure (or, as solid, liquid or frizzed fuel under conventional pressure). In this case the fuel density can reach $n = 10^{-26} \div 10^{-27}$ 1/m$^3$ (or solid, liquid, frozen fuel may be inside conductive matter, $n = 10^{-28} \div 10^{-29}$ 1/m$^3$). If the plasma reflectivity is high ($q > 0,99$), that is enough for thermonuclear ignition and keeping plasma under the radiation pressure and magnetic pressure. For current MCF the magnetic intensity is 5 T. For AB-Self-MCF the magnetic intensity may be about $10^4$ T. For AB-radiation reactor the radiation pressure is about $10^{10} \div 10^{13}$ N/m$^2$ (millions atm) (fig.15). We can neglect the outer magnetic force in AB-Reactor and we may design AB-Self-MCF reactor without very complex and expensive superconductivity magnetic system.

## Acknowledgement

The author wishes to acknowledge R.B. Cathcart for correcting the author's English and other useful advice.